\documentstyle[12pt,aps,epsfig]{revtex}
\begin{document}
\title{A mixed-mode shell-model theory for nuclear structure studies}
\author{\\V. G. Gueorguiev,\footnote{Corresponding author. Email address:
vesselin@phys.lsu.edu} W. E. Ormand,\footnote{Current address: Physics
Directorate, L-414, Lawrence Livermore national laboratory, P.O. Box 808,
Livermore, CA 94551} C. W. Johnson, and J. P. Draayer\\}
\address{{\it Department of Physics and Astronomy, Louisiana State 
University,}\\
{\it Baton Rouge, Louisiana 70803-4001}}

\maketitle
\vspace{0.5cm}
\begin{abstract}
We introduce a shell-model theory that combines traditional spherical states, which
yield a diagonal representation of the usual single-particle interaction, with
collective configurations that track deformations, and test the validity of this
mixed-mode, oblique basis shell-model scheme on $^{24}$Mg. The correct binding energy
(within 2\% of the full-space result) as well as low-energy configurations that have
greater than 90\% overlap with full-space results are obtained in a space that spans
less than 10\% of the full space. The results suggest that a mixed-mode shell-model
theory may be useful in situations where competing degrees of freedom dominate the
dynamics and full-space calculations are not feasible.
\end{abstract}


\vspace{1.5cm}

% 21.60.Cs Shell model
% 21.60.Ev Collective models
% 21.60.Fw Models based on group theory
% 27.30.+t 20<A<39

{Keywords: mixed-mode, symmetry-mixing, degrees of freedom, 
non-perturbative, \\
non-orthogonal basis, generalized eigenvalue problem}

{Numerical methods: Cholesky, Lanczos}

%\twocolumn
%\columnwith\csname@twocolumnfalse\endcsname

\section{Introduction}

Every quantum mechanical problem involves the solution of an eigenvalue equation.
In practice, there are only a few exactly solvable analytic models, and these are
realized only rarely (if ever) in nature so one is often forced to consider
numerical solutions. In addition, since in the latter case the dimensionality of a
model space can be very large, especially when dealing with realistic systems such
as nuclei where the number of particles is more than a few but less than what is
needed to justify the use of statistical methods, it is frequently necessary to
consider various approximations, including, especially, basis truncation.

The usual approach is to select a convenient orthonormal basis in which to carry
out a calculation. The use of a non-orthogonal scheme, though in principle no more
difficult than using an orthogonal one, is only justified if it is driven by
physical considerations.  For example, when a system supports competing modes and a
`preferred' basis can be associated with each, it makes sense to consider a
non-orthogonal basis comprised of leading configurations of each of these modes. 
Finding a `good' orthogonal basis, which is intimately related to a `good' (but not
exact) symmetries can be as difficult as solving the problem itself; nonetheless,
this is usually key to understanding the dominant modes (underlying physics) of a
system and outcomes of numerical calculations \cite{SCtdBS}.  

The mixed-mode system of interest to us seeks to accommodate the {\bf single-particle}
and {\bf collective-quadrupole} degrees of freedom that dominate the low-energy
structure of atomic nuclei \cite{2EssModes}. Manifestations of important collective
excitations can also be found in many other branches of physics \cite{CinPhys}. The
mixed-mode nature enters in nuclear physics because the single-particle and 
collective-quadrupole modes have similar excitation energies, both being small
relative to intra-shell excitation energies \cite{2EssModes,CinNP}. Both vibration
and rotation are manifestations of the  collective motion \cite{RandVbands_in_Nuclei},
vibrations being close to spherical shape oscillations and rotations being the periodic
motion of a well-deformed intrinsic configuration about some axis
\cite{Collectivitty_of_theRandVbands}. While in principle these modes are reachable
through multiple single-particle excitations, a full many-body theory is required
to offer a proper interpretation of this collective motion
\cite{CS_need_many_SPC,Elliott'sSU(3)}.

In this paper we explore the efficacy of a mixed-mode shell-model scheme by
considering $^{24}$Mg which is known to manifest strongly competing
single-particle and collective degrees of freedom. In particular, we examine
convergence of results towards those of full-space $0\hbar\omega$ $sd$-shell
calculations as a function of the number of particles in excited single-particle
levels and the number of irreducible representations (irreps) of SU(3). The
results show that a relatively small oblique space gives the right relative energy
for the K=2 band which leads to the correct order for all the low-energy levels.
The  structure of the states is further tested against the exact full-space
results by examining the overlap of calculated eigenstates. We begin by reviewing
the mathematical underpinning of the theory in the second section.  Results are
presented in the third section with conclusions and a discussion on applicability
of the approach given in the fourth section.

\section{Mathematical background}

The success of the current approach, which will be demonstrated in detail
in the next section, can be traced to the fact that the spherical shell-model
states are eigenstates of the one-body Hamiltonian ($\sum \varepsilon
_{i}a_{i}^{+}a_{i}$) while the two-body part of the Hamiltonian ($\sum_{i,j}
V_{kl,ij} a_{i}^{+}a_{j}^{+}a_{k}a_{l}$) is strongly correlated with $Q\cdot Q$
which is diagonal in the SU(3) basis \cite{QQ_in_sd-shell}.  By combining
spherical shell-model states and SU(3) states one accommodates, from the onset,
the dominant modes of the system.

The usual procedure for solving the eigenvalue problem $\hat{H}\vec{v} = \lambda
\vec{v}$  is to cast it into the form of a matrix equation. In a non-orthogonal
basis \cite{PSEVP-nonOrtogonal}, this matrix form includes the overlap matrix,
$\Theta_{ij}=\langle j|i \rangle$, and has the form $\sum_{j}[H_{ij}v_{j} - \lambda
\Theta_{ij}v_{j}] = 0$. For an orthonormal basis the overlap matrix becomes the
identity matrix ($\Theta_{ij}\rightarrow\delta_{ij}$) and the matrix form of the
eigenvalue problem is $\sum_{j} H_{ij}v_{j} = \lambda v_{i}$. When the overlap
matrix $\Theta$ is positive-definite the Cholesky algorithm can be used to solve the
generalized eigenvalue problem \cite{SnOEVP-Cholesky}. For large spaces this eigenvalue
problem can be solved efficiently by using the Lanczos algorithm
\cite{LanczosAlgorithm}.

For the calculations reported here we use a double mixed-mode basis.  The first set
consists of spherical shell-model states (ssm -- states) expressed in terms of
spherical single-particle coordinates ($nlj$). The second set has a good SU(3)
structure (su3 -- states) which track nuclear deformation \cite{SU3-(beta-gamma)};
this basis set is given in terms of cylindrical single-particle coordinates. By
construction both sets have the third  projection $M_{J}$ of the system's total
angular momentum J as a good quantum number \cite{m-scheme,good_Mj_and_SU(3)}. 
Schematically, these basis vectors and their overlap matrix can be represented in
the following way:

\begin{eqnarray}
{\rm basis\quad vectors} &:&{\rm \quad }\left(
\begin{array}{l}
e_{\alpha } :{\rm ssm\ - \ basis} \\
E_{i} :{\rm su3\ - \ basis}
\end{array}
\right),  \label{Basis vectors} \\
{\rm overlap\quad matrix} &:&{\rm \quad }\Theta =\left(
\begin{array}{ll}
{\bf 1} & \Omega \\
\Omega ^{+} & {\bf 1}
\end{array}
\right) ,\qquad \Omega _{\alpha i}=e_{\alpha }\cdot E_{i},
\label{Overlap matrix} \\
{\rm Hamiltonian\quad matrix} &:&{\rm \quad }H=\left(
\begin{array}{ll}
H_{ssm \times ssm} & H_{ssm \times su3} \\
H_{su3 \times ssm} & H_{su3 \times su3}
\end{array}
\right) =\left(
\begin{array}{ll}
H_{\alpha \beta } & H_{\alpha j} \\
H_{i\beta } & H_{ij}
\end{array}
\right) .
\label{hamiltonian Matrix}
\end{eqnarray}
In the above, $\alpha$ and $i$ span the following ranges: $\alpha
= 1$,  ..., dim(ssm -- basis) and $i = 1$, ..., dim(su3 -- basis).

Calculations in a nonorthogonal oblique-basis require an evaluation 
of the matrix elements of physical operators plus a knowledge of the scalar product
($e_{\alpha}\cdot E_{i}$). While it may be desirable to have an analytical
expression for the overlap matrix, as we have for the single-particle overlap
matrix \cite{AnaliticExpressionForM-Ovr}, for practical purposes it suffices
either to know the representation of each basis state in a common set that spans the full space, which is counter to the overall objective of reducing 
the number of basis states to a manageable subset, or to expand one set in terms of the
other.  For the present work, the $e_{\alpha}$, which can be represented by a
single machine word in a spherical single-particle scheme, were expanded in a
cylindrical basis, which is the representation for our collective SU(3)
basis vectors. This transformation is handled by an efficient routine that exploits
two computational aids:  bit manipulation via logical operations and a
weighted search tree for fast data storage and retrieval \cite{WST}. In the best
case scenario a transformation of this type has to be done at least once per ssm
basis state $e_{\alpha}$. We transform the ssm basis states since the result is
usually a vector with fewer components than a typical SU(3) basis state. There is
a simple way to calculate the overlap between states \cite{dot(a.b)=Det(A'B)},
however for the calculation of matrix elements it is better to transform each
$e_{\alpha}$ vector in the basis used by the SU(3) states.

Matrix elements of the one-body and two-body Hamiltonian
\begin{equation}
H=\sum_{i}\varepsilon _{i}a_{i}^{+}a_{i}+\frac{1}{4}
\sum_{i,j}V_{kl,ij}a_{i}^{+}a_{j}^{+}a_{k}a_{l} 
\label{Hamiltonian}
\end{equation}
have to be evaluated in each subspace ($H_{\alpha\beta}$ and $H_{ji}$) as well as
between the spaces ($H_{\alpha i}$ and $H_{j\beta }$), see (\ref{hamiltonian
Matrix}). The $H_{\alpha\beta}$ part is normally given and evaluated in a spherical
single-particle basis. By transforming the Hamiltonian to a cylindrical
single-particle basis one can obtain the $H_{ji}$ part of $H$. In order to compute
the off-diagonal blocks $H_{\alpha i}$ and $H_{j\beta}$ and overlap matrix elements
between SU(3) and ssm basis states, both basis sets are expanded in a basis of
Slater determinants using cylindrical single-particle states.  For the (8,4) and
(9,2) irreps this is an expansion into 2120 Slater determinants at most; each ssm
state, which itself is a single Slater determinant in a spherical single-particle
basis, typically expands into a smaller number of cylindrical-basis Slater
determinants which is less than 1296. We do not expand the SU(3) states into
spherical-basis Slater determinants because that would require significant fraction
of the entire spherical shell model space, defeating the rationale of our approach.
Taking into account the significant number of Hamiltonian matrix elements ($H_{ij}$
and $H_{i\beta}$) between multi-component states, it should be clear that this is
the most time consuming part of the calculation. The extra labels associated with
the intrinsic quadrupole moment $\varepsilon$ of each basis state is used to produce
well-structured band-like matrices and to speed up the calculation. Specifically,
basis states are pre-ordered according to their deformation as reflected by
$\varepsilon$ and during the evaluation of $H$ a $\Delta\varepsilon$ selection
rule  is applied.

It is important to point out that knowledge of the overlap matrix $\Theta $ and the
matrix elements of $H$ in the two spaces ($H_{\alpha \beta }$, $H_{ij}$) is not
enough to obtain the correct off-diagonal block $H_{\alpha i}.$  This is clear from
the following explicit expression for $H_{\alpha i}$ which contains a summation
along ($\bar{\beta}$) that lies outside of the ($\beta ,i$) model spaces:
\begin{equation} 
H_{\alpha i}=\sum_{\beta }H_{\alpha \beta }\Theta _{\beta i}+\sum_{\bar{
\beta }}H_{\alpha \bar{\beta}}\Theta _{\bar{\beta}i} .
\label{intermediate terms in H}
\end{equation} Thus a direct evaluation of $H_{\alpha i}$ is required.

It is instructive to consider a geometrical visualization of the oblique
basis-state concept. Since a set of vectors defines a hyper-plane, it 
is natural
to ask the question: ``What is the angle between hyperplanes defined by the
bases under consideration?'' To answer this question, first consider the angle
$\theta$ between a normalized SU(3) basis vector $\vec{v}$ and the 
subspace $V$ spanned by the spherical shell-model basis vectors. The length of the
projected vector $\vec{v}_{V}\in V$ is given by $\cos (\vec{v},V)=\cos \theta =\left|
\vec{v}_{V}\right|$. The space $V$ of the spherical shell-model basis vectors
induces a natural basis $\vec{n}_{\varepsilon }$ in the SU(3) space ($\vec{n}
_{\varepsilon }=n_{\varepsilon }^{i}\vec{E}_{i}$). The angle between each new
basis vector $\vec{n}_{\varepsilon }$ and the space $V$ will again be 
the length
of its projection into the space $V,$ but it has the nice property 
that this set
of orthogonal basis vectors stays orthogonal after the projection 
into the space
$V$:
\begin{eqnarray*}
\cos \theta _{\varepsilon } &=&\cos (\vec{n}_{\varepsilon },V)=\left| \vec{n}
_{\varepsilon V}\right|, \\
\vec{n}_{\varepsilon V} &=&\sum_{i,\alpha }n_{\varepsilon }^{i}(\vec{E}%
_{i}\cdot \vec{e}_{\alpha })\vec{e}_{\alpha }=\sum_{i,\alpha }n_{\varepsilon
}^{i}\Theta _{i\alpha }\vec{e}_{\alpha }, \\
\left| \vec{n}_{\varepsilon V}\right| ^{2} &=&\sum_{\alpha
}(\sum_{i}n_{\varepsilon }^{i}\Theta _{i\alpha })^{2}=\sum_{\alpha
,i,j}n_{\varepsilon }^{i}\Theta _{i\alpha }n_{\varepsilon }^{j}\Theta
_{j\alpha }.
\end{eqnarray*}
In matrix notation this reads
\[
\left| \vec{n}_{\varepsilon V}\right| ^{2}=\vec{n}_{\varepsilon }\cdot \hat{
\Theta }\cdot \hat{\Theta}^{T}\cdot \vec{n}_{\varepsilon },
\]
where the natural basis vectors $\vec{n}_{\varepsilon }$ are eigenvectors of
the symmetric matrix $\hat{\Theta}\cdot \hat{\Theta}^{T}$
\begin{equation}
\hat{\Theta}\cdot \hat{\Theta}^{T}\cdot \vec{n}_{\varepsilon }=\varepsilon
^{2}\vec{n} _{\varepsilon }.  \label{natural basis}
\end{equation}
It follows that $\left| \vec{n}_{\varepsilon V}\right| ^{2}=\vec{n}
_{\varepsilon }\cdot \hat{\Theta}\cdot \hat{\Theta}^{T}\cdot 
\vec{n}_{\varepsilon
}=\varepsilon ^{2}\vec{n}_{\varepsilon }\cdot \vec{n}_{\varepsilon
}=\varepsilon ^{2}$ and thus the matrix $\hat{\Theta}\cdot
\hat{\Theta} ^{T}$ is positive definite ($\left| \vec{n}_{\varepsilon V}\right|
^{2}=\varepsilon ^{2}\geq 0$ ) with eigenvalues determined by the 
$\cos \theta .$
This construction allows for a simple visualization of the oblique basis space:
Choose the $x-$axis to correspond to the space $V$ of all the spherical
shell-model basis vectors and represent the SU(3) space as a 
collection of unit
vectors each at an angle $\cos \theta =\varepsilon $ with respect to the
$x-$axis. In the next section this construction will be applied to the geometry
of oblique basis space calculations to demonstrate the relative orthogonality
of the two vector sets, $e_{\alpha}$ and $E_i$.

\section{An illustrative example}

In this section the oblique-basis technique is tested for $^{24}$Mg, which is a
strongly deformed nucleus with well-known collective properties and one of the
best manifestations of the Elliott's SU(3) symmetry
\cite{Elliott'sSU(3)}. In terms of dimensionality of the model space, 
adding a few leading SU(3) irreps to a highly truncated spherical shell-model 
basis results in significant gains in the convergence of the low-energy spectra 
towards the full space result. In particular, the addition of leading SU(3)
irreps yields the right placement of the K=2 band and the correct order for
most of the low-lying levels. Indeed, an even more detailed analysis shows that 
the structure of the low-lying states is significantly improved through the
addition of a few SU(3) irreps. The Hamiltonian used in our analysis is the
Wildenthal interaction \cite{Wildenthal}.

Our model space for $^{24}$Mg consists of 4 valence protons and 4 valence neutrons
in the $0\hbar\omega$ $sd$-shell. The $m$-scheme dimensionality ($M_{J}=0$) of this
space is 28503 which in the figures that follow is denoted FULL. To test the
effects of truncations, calculations were also carried out on permitting $n$ particles
to be excited out of the lowest $d_{5/2}$ orbit, i.e.
$d_{5/2}^{8-n}(d_{3/2}s_{1/2})^n$, and are denoted as SM(n). The SM(2) approximation is
of particular interest since it allows one to take into account the effect of pairing
correlations (one pair max) in the `secondary levels' ($s_{1/2}$ and $d_{3/2}$ for the
ds shell) with a minimum expansion of the model space. The SU(3) part of the basis
includes two scenarios: one with only the leading representation of SU(3) included,
which for $^{24}$Mg is the (8,4) irrep with dimensionality 23 for the $M_{J}=0$
space and denoted in what follows by appending (8,4) to the corresponding SM(n)
notation; and another with the (8,4) irrep plus the next most important
representation of SU(3), namely the (9,2). The (9,2) irrep occurs three times,
once with S=0 ($M_{J}=0$ dimensionality 15) and twice with S=1 ( $M_{J}=0$
dimensionality $2\times 45=90$). All three (9,2) irreps have total $M_{J}=0$
dimensionality of 15+90=105. The (8,4)\&(9,2) case has total $M_{J}=0$
dimensionality of 23+105=128 and is denoted by appending (8,4)\&(9,2) to the
corresponding SM(n) notation. In Table \ref{Table1} we summarize the dimensionalities
involved.

\vspace{0.5cm}

\begin{table}[tbph]
\caption{Labels used to distinguish various calculations and the corresponding
$M_{J}=0$ dimensionalities. The leading SU(3) irrep is denoted by (8,4) while
(8,4)\&(9,2) implies that (9,2) irreps have also been included. The SM(n) spaces
correspond to spherical shell-model partitions with $n$ valence particles excited out
of the $d_{5/2}$ shell into the $s_{1/2}$ and
$d_{3/2}$ levels.}
\vskip 0.25cm
\begin{tabular}{rrrrrrrrrrrrrrr}
Model space &  & $(8,4)$ &  & $(8,4)\&(9,2)$ & & $SM(0)$ &
& $SM(1)$ &  & $SM(2)$ &  & $SM(4)$ &  & $FULL$ \\ \hline
space dimension &  & $23$ &  & $128$ & & $29$ &  & $449$ &  &
$2829$ &  & $18290$ &  & $28503$ \\
\% of the full space &  & $0.08$ &  & $0.45$ & & $0.10$ &  & $
1.57$ &  & $9.92$ &  & $64.17$ &  & $100$
\end{tabular}
\label{Table1}
\end{table}

\begin{figure}[tbph]
\centerline{\hbox{\epsfig{figure=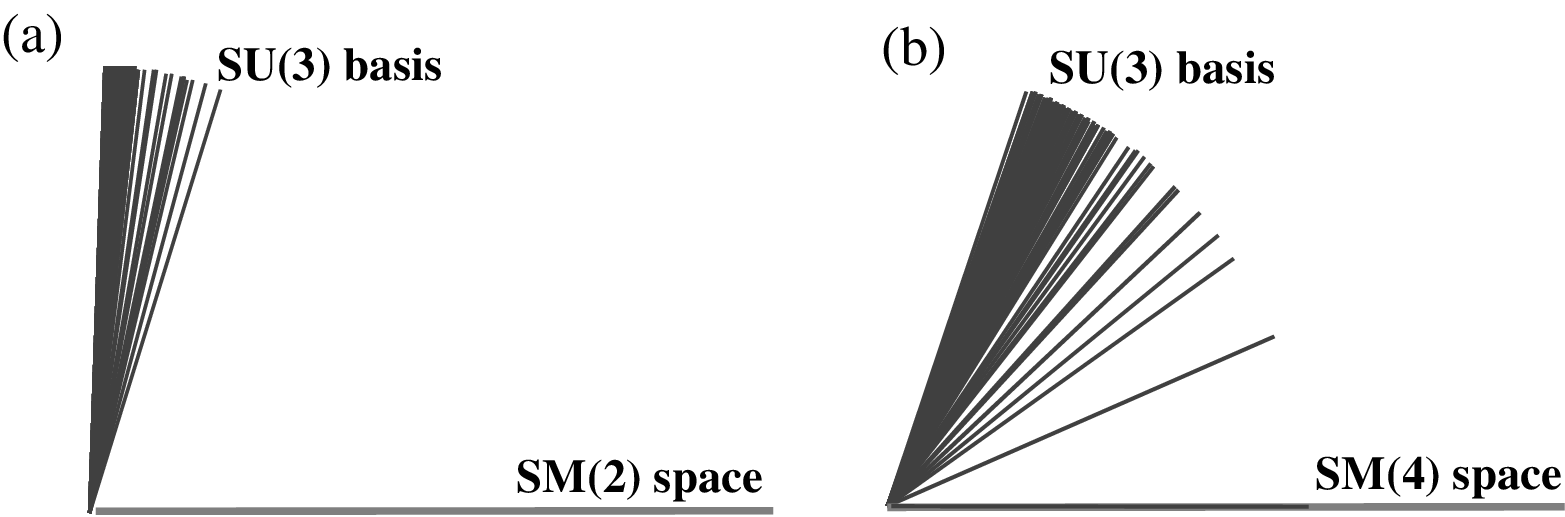,width=16cm,height=5cm}}}
\vskip 0.5cm
\caption{Orthogonality of the basis vectors in the oblique geometry.
The SU(3) space consist of (8,4)\&(9,2) basis vectors with the
shell-model spaces (SM(n) with n=2 and 4) indicated by a horizontal
line. (a) SM(2) and the natural SU(3) basis vectors and (b) SM(4) and
the natural SU(3) basis vectors. In the latter case there
are five SU(3) vectors that lie in the SM(4) space.}
\label{SU3+ Relative To SM(2) and SM(4)}
\end{figure}

Now, the method described at the end of the previous section is used to visualize
the structure of the oblique basis space. First consider the SM(2) space enhanced
by the SU(3) irreps (8,4)\&(9,2). Since the SM(2) and (8,4)\&(9,2)
spaces are both relatively small (see Table \ref{Table1}) we expect the basis
vectors in these spaces to be nearly orthogonal. This orthogonality is clearly seen
from inset (a) in Fig.\ref{SU3+ Relative To SM(2) and SM(4)}. Inset (b) in
Fig.\ref{SU3+ Relative To SM(2) and SM(4)} shows a loss of orthogonality between the
SM(4) and the (8,4)\&(9,2) basis vectors. This is due to the fact that SM(4)
space is about 64\% of the full $sd$-space and therefore there is a relatively
high probability that some linear combinations of the SU(3) basis vectors lie in
the SM(4) space. Indeed, it can be shown that there are five vectors from
(8,4)\&(9,2) that lie within the SM(4) space. Such redundant vectors must, of
course, be excluded from the calculation.

%use as a model
%\centerline{\hbox{\epsfig{figure=SU3SM2SM4.eps,width=14cm,height=10cm}}}

\begin{figure}[tbph]
\centerline{\hbox{\epsfig{figure=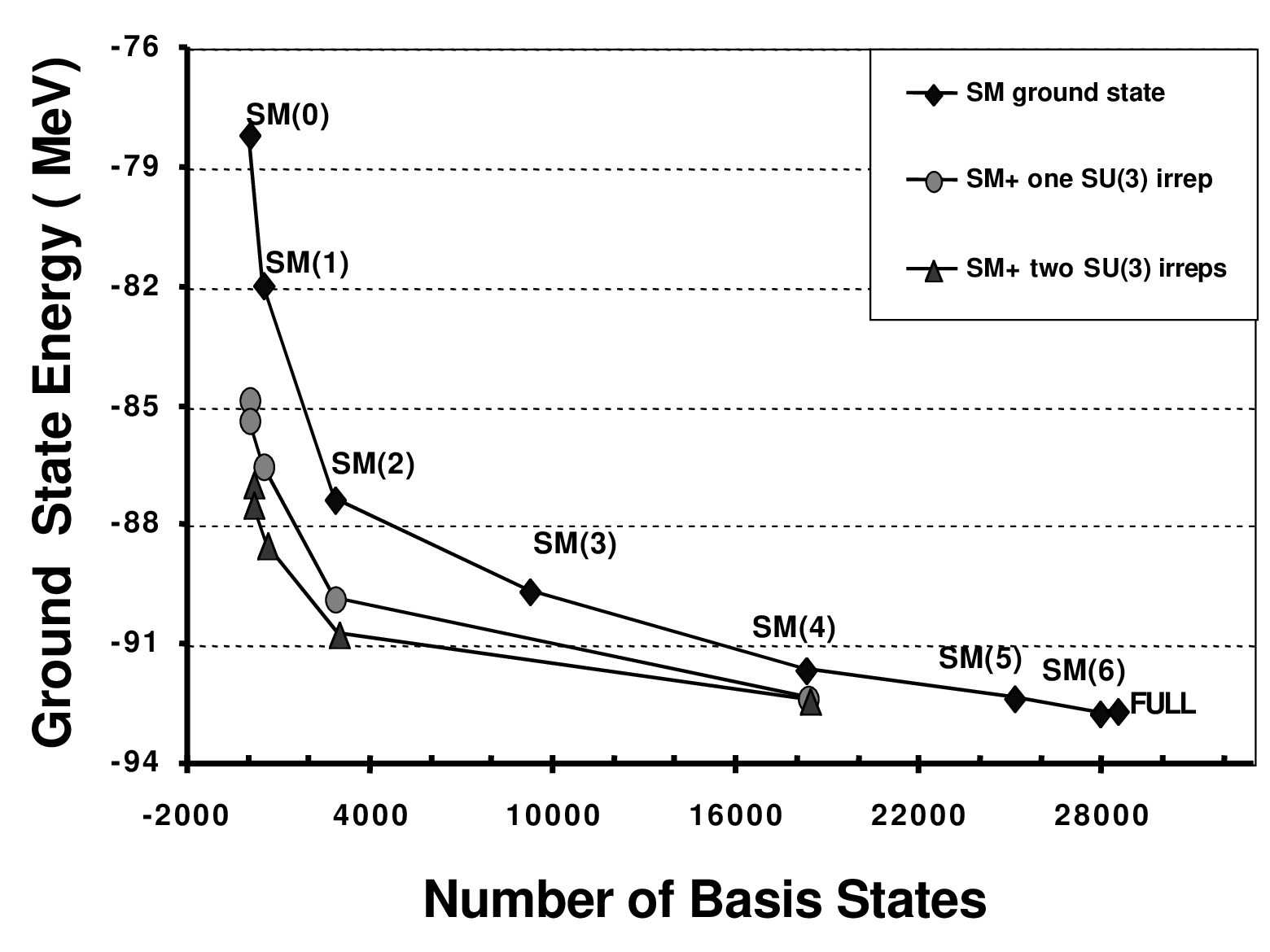,width=16cm,height=12cm}}}
\caption{Calculated ground-state energy for $^{24}$Mg as a function of the various
model spaces. Note the dramatic increase in binding (3.3 MeV) in going from SM(2) to
SM(2)+(8,4)\&(9,2) (a 0.5\% increase in the dimensionality of the model space).
Enlarging the space from SM(2) to SM(4) (a 54\% increase in the dimensionality of the
model space) adds 4.2 MeV in binding energy.}
\label{Mg24DimConv}
\end{figure}

We now turn to a consideration of the main results of the oblique basis calculation,
starting with ground-state convergence issues. The  results shown in
Fig.\ref{Mg24DimConv} illustrate that the oblique basis calculation gives good
dimensional convergence in the sense that the calculated ground-state  energy for the
SM(2)+(8,4)\&(9,2) calculation is 3.3 MeV below that  calculated energy for the SM(2)
space alone. Adding the SU(3) irreps only increases the size of the space from 9.9\% to
10.4\% of the full space. This 0.5\% increase in the size of the space is to be
compared with the huge (54\%) increase  in going from SM(2) to a SM(4) calculation.
For the later the ground state energy is 4.2 MeV lower than SM(2) result, somewhat
better than for the SM(2)+(8,4)\&(9,2) calculation but in 64.2\% rather than 10.4\% of
the full model space.

%use as a model
%\centerline{\hbox{\epsfig{figure=Mg24DimConv.eps,width=16cm,height=12cm}}}

\begin{figure}[tbph]
\centerline{\hbox{\epsfig{figure=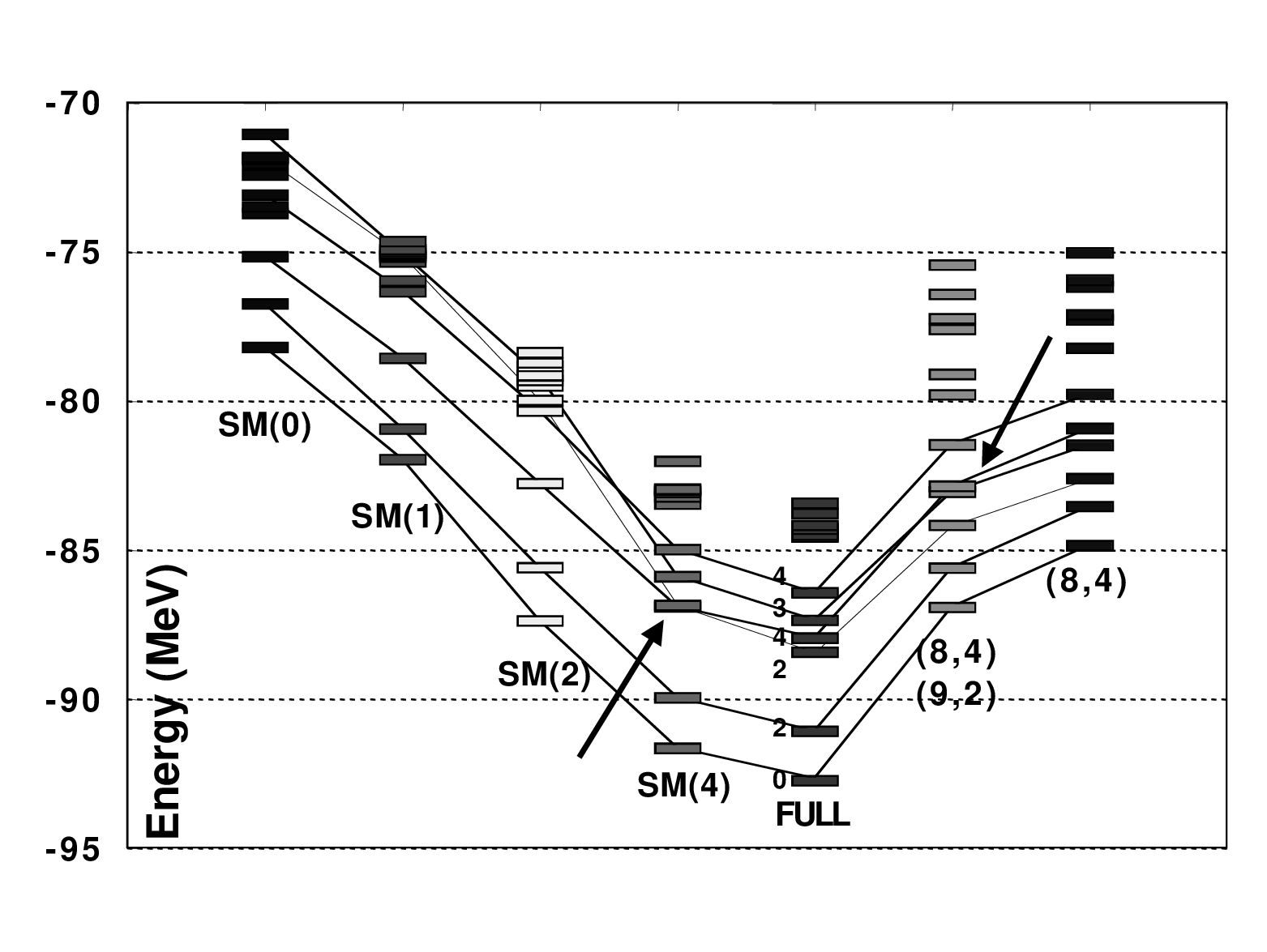,width=16cm,height=12cm}}}
\caption{Structure of the energy levels for $^{24}$Mg: pure 
$m$-scheme spherical
basis space calculations are on the left-hand side of the graph; pure 
SU(3) basis
space calculations are on the right-hand side; the spectrum from the 
FULL space
calculation is in the center.}
\label{LevelStructure}
\end{figure}

\begin{figure}[tbph]
\centerline{\hbox{\epsfig{figure=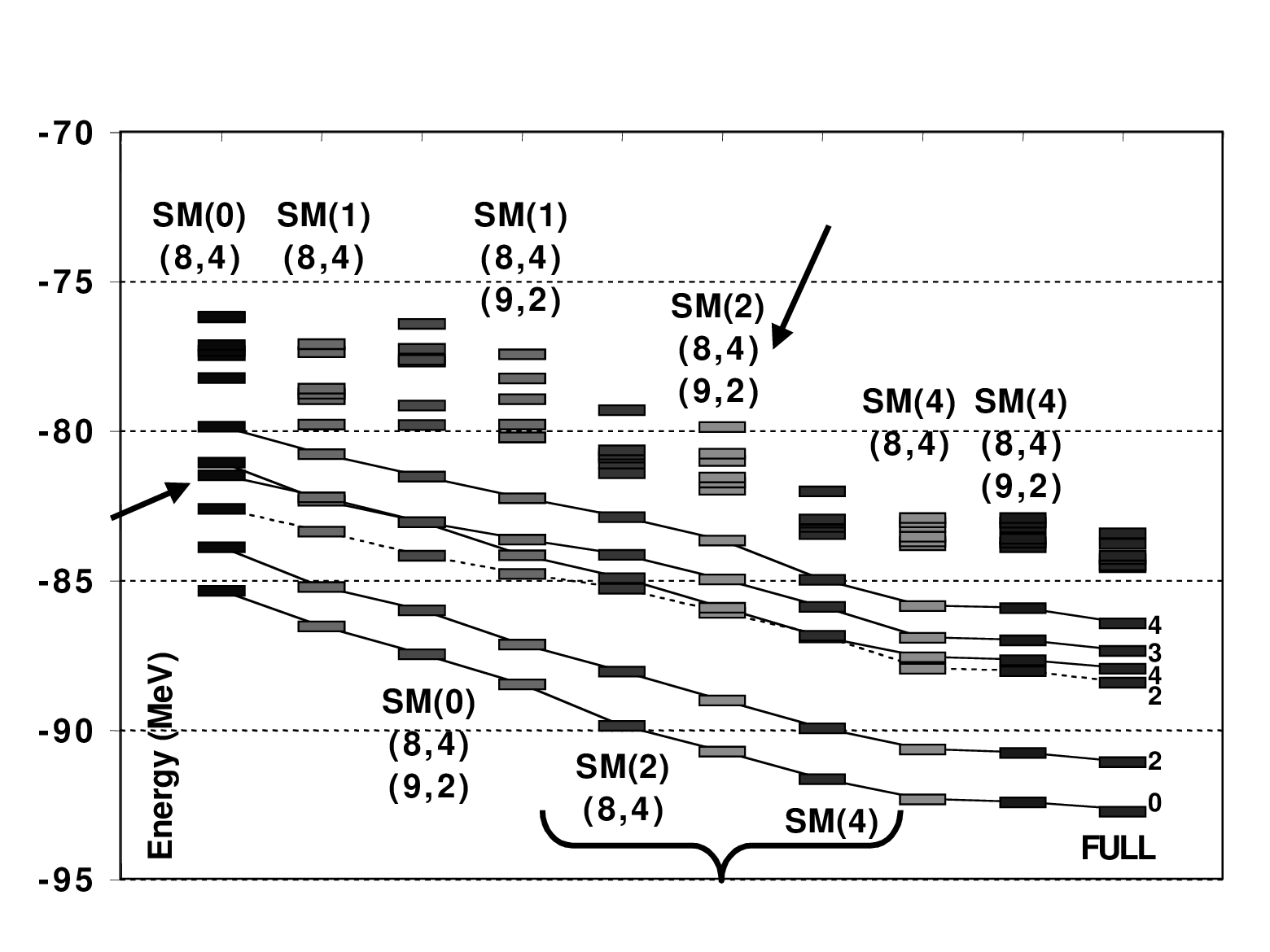,width=16cm,height=12cm}}}
\caption{Energy levels for $^{24}$Mg as calculated for different oblique bases.
The SM(4) basis calculation is included for comparison.}
\label{RightLevelStructure}
\end{figure}

Figs. \ref{LevelStructure} and \ref{RightLevelStructure} show that the oblique
basis calculation positions the K=2 band head correctly. Furthermore, most of the
other low-energy levels are also positioned correctly. The results for pure
spherical and pure SU(3) calculations are shown in Fig.\ref{LevelStructure}. As
can be see from the Fig.\ref {LevelStructure} results, an SM(4) calculation
(64\% of the full model space) is needed to get the ordering of the lowest angular
momentum states correct. Also notice that in this case the 3th and 4th energy
levels are practically degenerate. On other hand, it only takes 0.5\% of the full
space to achieve comparable success with SU(3). In particular, Fig.\ref
{LevelStructure} shows that an SU(3) calculation using only the (8,4) and
(9,2) irreps gives the right ordering of the lowest levels. Note that the first
few low-energy levels for SM(2) are close in energy to the corresponding
low-energy levels for the (8,4)\&(9,2) result. Since these two  spaces are nearly
orthogonal (see Fig.\ref{SU3+ Relative To SM(2) and SM(4)}), these two sets of
levels mix strongly in an oblique calculation and yield excellent results. The
comparable ground-state energies of the SM(2) and (8,4)\&(9,2) configurations
can also be seen from Fig.\ref{Mg24DimConv}.

%use as a model
%\centerline{\hbox{\epsfig{figure=LevelStructure.eps,width=16cm,height=12cm}}}

The spectra shown in Fig.\ref{LevelStructure} are to be compared with the results from
the oblique basis calculations shown in Fig.\ref{RightLevelStructure}. From this
comparison one can see that  the correct level structure can be achieved by using 1.6\%
(SM(1)+(8,4)) of the full $sd$-space. However, one should also notice that for the
SM(0)+(8,2) space, which is only 0.2\% of the full space and the minimum oblique basis
calculation, the results are quite close to the correct level structure. Despite the
fact that the ground state energy of the oblique basis calculations is higher than the
ground state energy for the SM(4) type calculation, the oblique calculations are
favorable in terms of dimensionality considerations and correctness of the level
structure.

%use as a model
%\centerline{\hbox{\epsfig{figure=RightLevelStructure.eps,width=16cm,height=12cm}}}

\begin{figure}[tbph]
\centerline{\hbox{\epsfig{figure=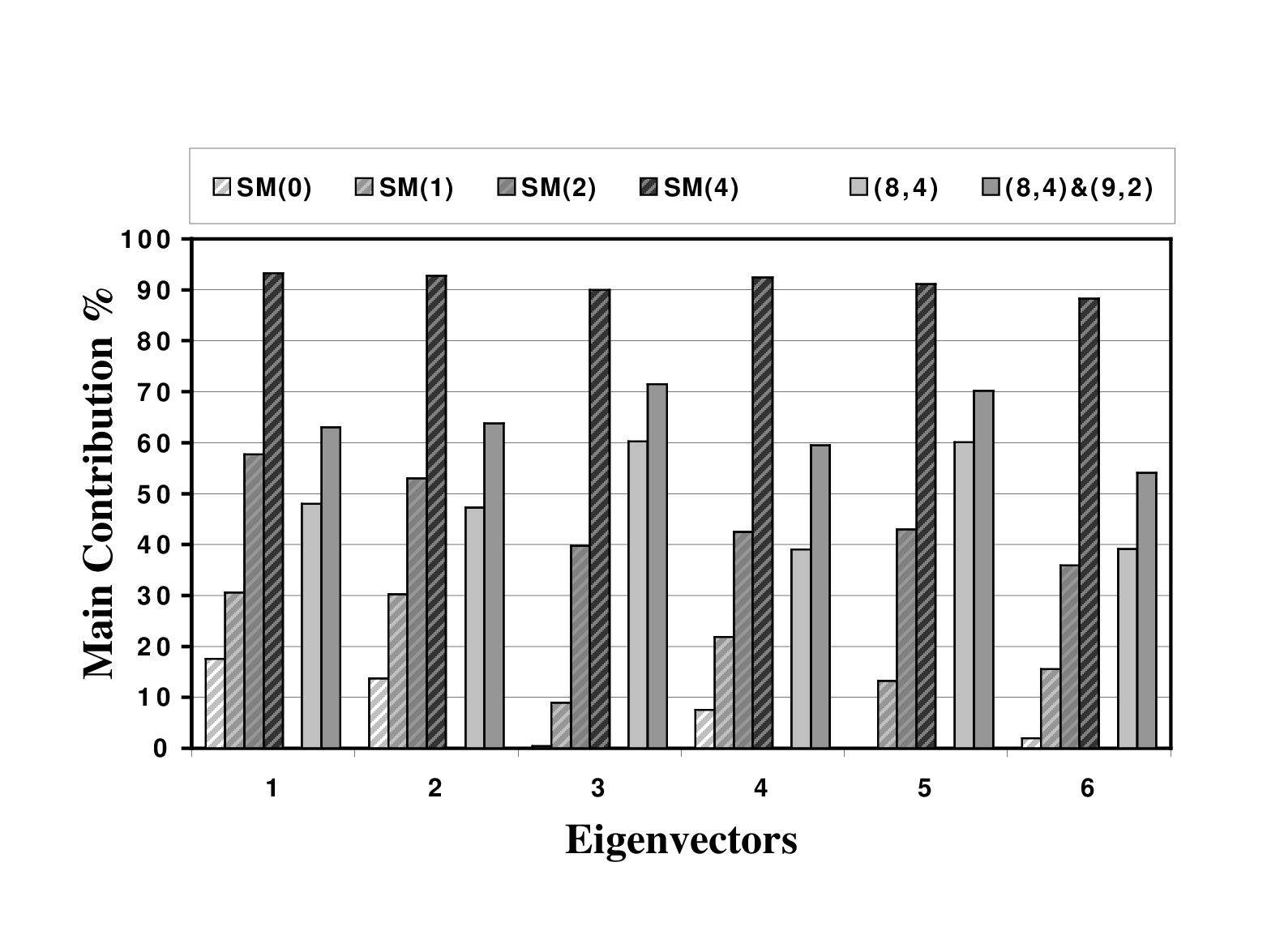,width=16cm,height=12cm}}}
\caption{Overlaps of the pure spherical shell-model and pure SU(3) 
eigenstates
with the corresponding FULL space results for $^{24}$Mg. The first four bars
represent the SM(0), SM(1), SM(2), and SM(4) calculations, the
next three bars represent SU(3) calculations, etc.}
\label{UsualCalculationOverlaps}
\end{figure}

\begin{figure}[tbph]
\centerline{\hbox{\epsfig{figure=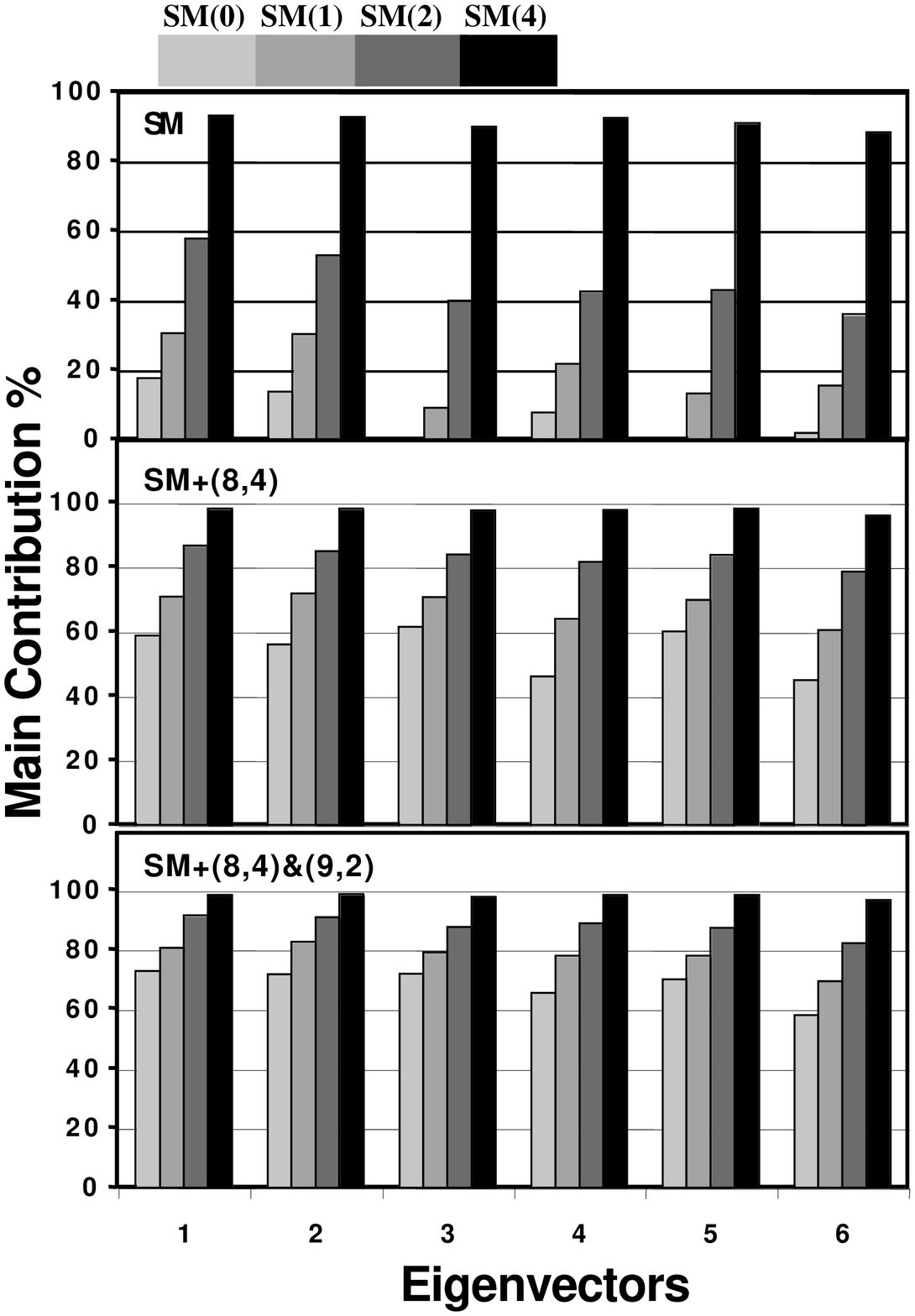,width=16cm,height=20cm}}}
\caption{Overlaps of calculated eigenstates for oblique basis calculations with the
exact results from FULL $sd$-shell calculations. Inset SM contains the overlaps for the
pure spherical shell-model basis states  only. Inset SM+(8,4) contains the overlaps of
the SM basis enhanced by the leading SU(3) irrep (8,4). Inset SM+(8,4)\&(9,2) has the
(9,2) irreps included as well.}
\label{MixedOverlaps}
\end{figure}

\begin{figure}[tbph]
\centerline{\hbox{\epsfig{figure=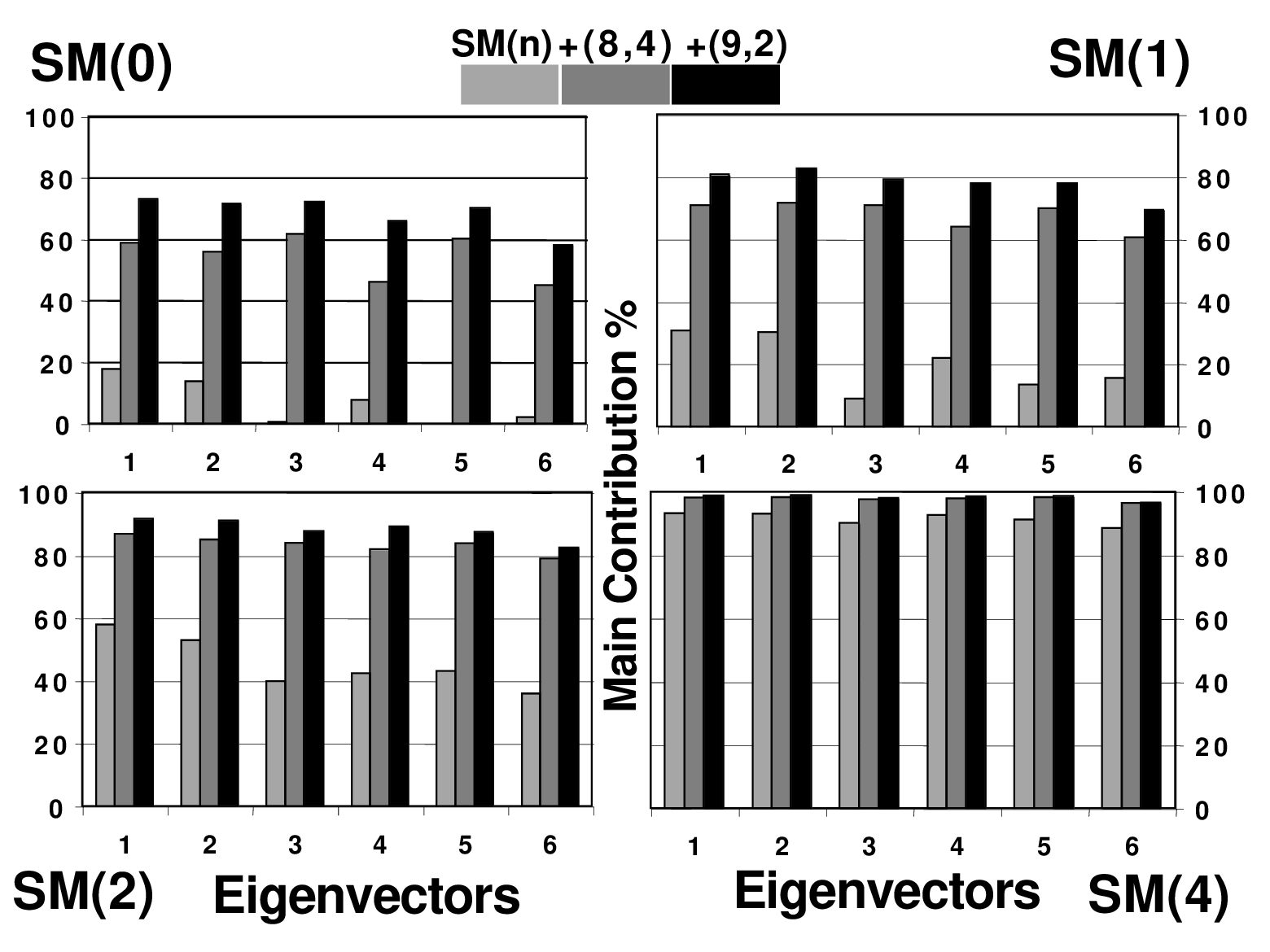,width=16cm,height=12cm}}}
\caption{Overlaps of oblique basis states with the exact eigenstates from
the FULL $sd$-shell calculation. Each inset represents a particular
SM(n)--type calculation, showing how the overlaps change along the 
corresponding
oblique basis calculation.}
\label{MixedCalculationOverlaps}
\end{figure}

\begin{figure}[tbp]
\centerline{\hbox{\epsfig{figure=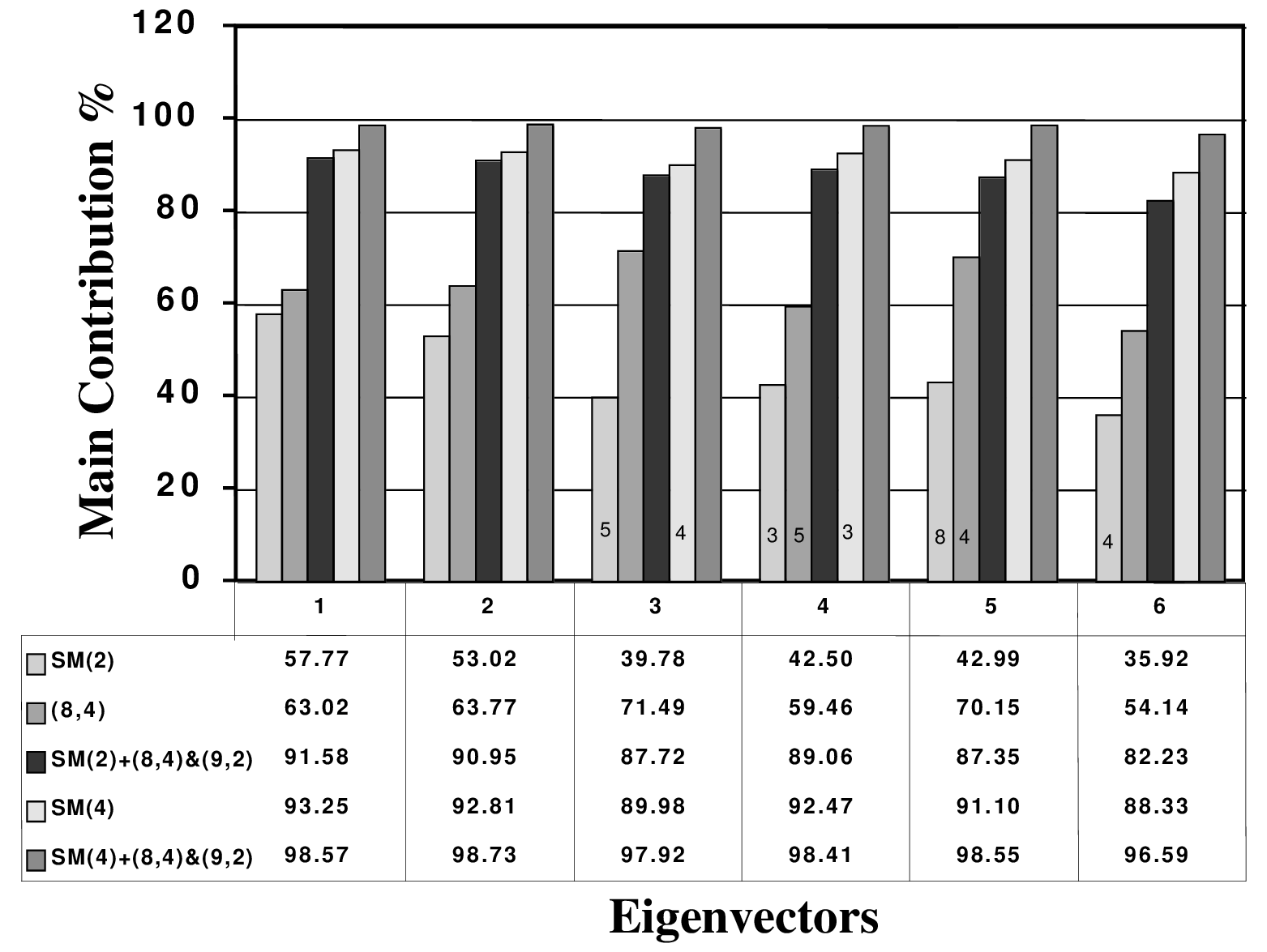,width=16cm,height=12cm}}}
\caption{Representative overlaps of pure SM(n), pure SU(3), and oblique
basis results with the exact full $sd$-shell eigenstates. Numbers 
within some of the bars reflect the miss-ordering of the corresponding spectrum with 
respect to the exact result. The number points to the actual state that has
the overlap shown by the bar, this is the maximal overlap observed.}
\label{SelectedOverlaps}
\end{figure}

Figs. \ref{UsualCalculationOverlaps}--\ref{SelectedOverlaps} focus on the
actual structure of the states by showing overlaps of eigenstates calculated
in the SM(n), SU(3), and oblique bases with the corresponding states of the
full space calculation. Specifically, in Fig.\ref{UsualCalculationOverlaps}
overlaps of states for pure SM(n) and pure SU(3) type 
calculations are given. Note that the SM(4) states have big overlap (90\%) for
the first few eigenstates. This should not be too surprising since SM(4) covers 
64\% of the
full space.

%use as a model
%\centerline{\hbox{\epsfig{figure=UsualCalcOverlaps.eps,width=16cm,height=12cm}}}

The results of Fig.\ref{UsualCalculationOverlaps} show that in general SU(3) based
calculations give much better results then low-dimensional SM(n)-type calculations. The
SM(n) based calculations have irregular overlaps along the low-lying states and require
SM(4), which is 64\% of the full space, to get relatively well behaved overlaps. This
can be seen most clearly from the inset labeled SM in Fig.\ref{MixedOverlaps}. Note
that the SM(0) contributions to the third, fifth and sixth states are very low, while
SM(1) and SM(2) have varying contributions. The structure of states obtained in any
SU(3)--type calculation leads to a stable picture for the oblique calculations as shown
in the inset SM(n)+(8,4) and SM(n)+(8,4)\&(9,2) in Fig.\ref{MixedOverlaps}.

%use as a model
%\centerline{\hbox{\epsfig{figure=MixedOverlaps.eps,width=16cm,height=20cm}}}

In Fig.\ref{MixedCalculationOverlaps} the improvement in the structure of
the calculated states is followed as the SU(3) states are added to the
SM(n) basis. From this graph one can see that the improvement to the
SM(0)-- and SM(1)-type calculation is due mainly to the goodness of SU(3)
itself. The improvement obtained in the oblique calculation is due to 
the SU(3) enhancement of the SM(2) space. From this graph one can also conclude
that there is only a small gain in going to the SM(4) based oblique calculation.
However, this improvement can not be achieved by any other means with such a
small increase in the model space. This is clear from a careful examination of
Fig.\ref{Mg24DimConv} where one can see that the SM(5) result, which has
25142 basis vectors (88\% of the full $sd$-space), gives the same 
ground-state energy as the SM(4)+(8,4)\&(9,2) result (64.6\% of the full
$sd$-space).

%use as a model
%\centerline{\hbox{\epsfig{figure=MixedCalcOverlaps.eps,width=16cm,height=12cm}}}

Finally, to compare the three schemes -- SU(3), SM(n) and the various oblique basis
combinations -- representative overlaps are shown in Fig.\ref {SelectedOverlaps}. From
these results it is very clear that SU(3)-type basis states yield the right structure
in very low order. In particular, in Fig.\ref{SelectedOverlaps} it can be seen that a
90\% overlap with the exact eigenvectors can be achieved by using only 10\% of the
total space, SM(2)+(8,4)\&(9,2). Furthermore, Fig.\ref{SelectedOverlaps} also shows that
SU(3) enhances the SM(4) results yielding eigenstates with  overlaps that are very
close ( $\approx$ 98\%) to the exact results.

%use as a model
%\centerline{\hbox{\epsfig{figure=SelectedOverlaps.eps,width=16cm,height=12cm}}}

\section{Conclusion and Discussions}

In this paper we have shown that knowledge about the important modes of a
physical system can be used to obtain good eigenstates in relatively small
model spaces. In particular, for the $^{24}$Mg example with typical
one-body and two-body interactions, the proposed oblique scheme gives good
dimensional convergence for the ground state energy, good spectral structure
for all the low-energy states, and good overlaps of these states with full
$sd$-shell calculations.

There are some natural choices for further development of the theory and its
application. The most straightforward is a study of other $sd$-shell nuclei as
well as $pf$-shell nuclei. Such studies will further test the theory 
and the codes that have been develop. These are currently underway. Another
possibility is to integrate the oblique basis concept into no-core calculations
of the type developed by \cite{Navratil'00}. Such an extension would involve
the symplectic group for multi-shell correlations rather than just SU(3)
\cite{Sp(6)models}. A third even broader extension of the theory would involve
a general procedure for the identification of dominant modes from any one- and
two-body Hamiltonian along with a complementary partitioning of the  model
space into physically relevant subspaces with small overlaps. One can then
start with eigenstates for an arbitrary  subspace and constructively improve
the results by including corrections from the remaining subspaces. It should be
possible to do this by keeping only a small set of the calculated lowest energy
states at each iteration.

The results presented here show very clearly that when important modes can
be isolated one can build an oblique theory that incorporates leading
configurations of each mode and get good convergence in a limited model space.

\vskip .5cm

Support provided by the Department of Energy under Grant No. DE-FG02-96ER40985,
and the National Science Foundation under Grant No. PHY-9970769, and by
Cooperative Agreement EPS-9720652 that includes matching from the 
Louisiana Board
of Regents Support Fund.

\end{document}